\documentclass{sf2a-conf2015}
\usepackage{graphicx}
\usepackage{hyperref}
\usepackage[]{natbib}
\usepackage{bm,bbm}
\usepackage[lined,ruled,vlined,linesnumbered]{algorithm2e}  
\usepackage{epstopdf}
\usepackage{amsmath}

\def\BibTeX{{\rm B\kern-.05em{\sc i\kern-.025em b}\kern-.08em
    T\kern-.1667em\lower.7ex\hbox{E}\kern-.125emX}}
\bibpunct{(}{)}{;}{a}{}{,}  

\DeclareMathOperator{\ST}{ST}

\DeclareMathOperator{\proj}{Proj}

\def\x{{\mathbf x}}
\def\V{{\mathbf V}}
\def\M{{\mathbf M}}
\def\F{{\mathbf F}}
\def\W{{\mathbf W}}

\begin{document}

\TitreGlobal{SF2A 2015}


\title{Interferometric Radio Transient Reconstruction in Compressed Sensing Framework}

\runningtitle{Interferometric Radio Transient Reconstruction}

\author{M. Jiang}\address{Service d'Astrophysique, CEA Saclay, Orme des Merisiers, 91410 GIF-Sur-YVETTE, France}

\author{J. N. Girard$^1$}

\author{J.-L. Starck$^1$}

\author{S. Corbel$^1$}

\author{C. Tasse}\address{GEPI, Observatoire de Paris-Meudon, 5 rue place Jules Janssen, 92190 Meudon, France}



\setcounter{page}{237}


\maketitle


\begin{abstract}

Imaging by aperture synthesis from interferometric data is a well-known, but is a strong ill-posed inverse problem. Strong and faint radio sources can be imaged unambiguously using time and frequency integration to gather more Fourier samples of the sky. However, these imagers assumes a steady sky and the complexity of the problem increases when transients radio sources are also present in the data. Hopefully, in the context of transient imaging, the spatial and temporal information are separable which enable extension of an imager fit for a steady sky. We introduce independent spatial and temporal wavelet dictionaries to sparsely represent the transient in both spatial domain and temporal domain. These dictionaries intervenes in a new reconstruction method developed in the Compressed Sensing (CS) framework and using a primal-dual splitting algorithm. According to the preliminary tests in different noise regimes, this new ``Time-agile'' (or 2D-1D) method seems to be efficient in detecting and reconstructing the transients temporal dependence.
\end{abstract}



\section{Introduction}
\label{sec:intro}
The study of the radio sky at radio wavelengths has increased since the arrival of new sensitive instrumentation. The timescale to produce images in radio was reduced by the development of new imaging techniques taking the full advantage of the instrument. At the same time, the study of known class of transient sources (e.g. pulsars for general relativity tests, Active Galactic Nuclei, etc) and the recent discovery of new class of transients (e.g. Rotating Radio Transients, Fast Radio Bursts, Lorimer type bursts, see \cite{Lorimer_2007}) has motivated further development for transient detection and characterization.

Imaging via aperture synthesis with interferometric data has been an active field of research for $\sim$40 years.
A radio interferometer gives a limited set of noisy Fourier samples of the sky (the visibilities \citep{wilson09}) inside the field of view of the instrument. An approximate of the sky can be obtained (assuming the small field approximation) by simply taking the inverse Fourier Transform (FT) of those visibilities. Due to the limited number of baselines, the Fourier plane is incomplete and one requires to process this incomplete Fourier map either by solving a deconvolution problem, using tools such as CLEAN and its derivates (e.g. \citet{clean,cleancs,MSMFS}) or by solving the ``inpainting'' problem by recovering missing information in the Fourier plane. Several teams have addressed this issue within the framework of the recent Compressed Sensing (CS) theory (e.g. \citet{Garsden_2015,Girard_2015,Dabbech2015} and other references therein).
In addition, next generation of giant interferometers such as LOFAR \citep{haarlem}, suffers from ``direction-dependent'' effects \citep{tasse2012} which distort the Fourier relationship between the measurements and the sky (such as array non-coplanarity and dipole projection). In \citet{Garsden_2015}, a new imager compatible with LOFAR combined both a sparse approach given by the CS theory and corrections for A and W effects \citep{tasse13}. It also demonstrated better angular resolution and lower residuals as compared to classical methods, on simulated and real datasets.

A lot of effort has been put into the development of detection pipelines (e.g. the LOFAR TRAnsient Pipeline -- TRAP \citep{Swinbank_2015}, based on fast iterative closed-loop performing calibration / imaging / source detection / catalogue cross-matching). However, being variable and mostly point-like, the transients imaging suffers from the imaging rate. On the one hand, short time integration enables temporal monitoring of a transient, but each snapshot provides poor visibility coverage. Therefore, the image has low signal-to-noise ratio (SNR) due to large fraction of missing data. On the other hand, long time integration ensures a good sampling, but it will average out the temporal variability of the transient by mixing and diluting ``ON'' state periods with ``OFF'' state periods.  As a result, a variety of transient radio sources might be missed due to uncertainty or timescale filtering. Consequently, it is difficult to use classical imagers to detect and image transient source when the temporal variability of the transient source is unknown. There is an interest in developing fast imagers, enable to cope with the time variability of the sources. Such imagers can rely on the CS framework to give a quick approximate of the true sky, giving access to faster transients. This motivated the development of a 2D-1D sparse reconstruction imager on the experience obtained in the 2D imaging case.
  
\section{Radio transient reconstruction in compressed sensing framework}
\label{sec:radioRec}

\subsection{Compressed sensing theory}
\label{ssec:CS}
Shannon theory is commonly known in the domain of signal processing to perfectly reconstruct a regularly sampled signal. However, the innovative sampling and compression theory of recent years, Compressed Sensing (CS) \citep{Candes2008}, or Compressive Sensing could go beyond the Shannon limit, at a rate significantly below the Nyquist rate, to capture and represent compressible signals based on the sparsity of observed signals.

The CS theory is a paradigm for finding a nearly exact reconstruction in the case of an undetermined problem. In the radio interferometry imaging problem, as we have fewer observations than unknowns (i.e. the sparsely sampled FT of the sky), the CS theory applies and could enable to produce accurate maps possibly with improved angular resolution. To achieve the perfect reconstruction from few samples, the CS theory relies on two principles: sparsity and incoherence. First, in general, the CS theory exploits the fact that the signal can be sparse or compressive in some dictionary $\mathbf{\Phi}$. For instance, a signal $\x(t)$ may be not sparse in the direct space, but can be sparse in the wavelet space. In such case, $\x(t)$ can be decomposed as its sum of few, but significant, coefficients, as $\x=\mathbf{\Phi}\bm{\alpha}=\sum_{i=1}^T \alpha[i]\varphi_{i}$, with T relatively small. Second, the incoherence principle states that a sparse signal in the dictionary $\mathbf{\Phi}$ must be as dense as possible in the domain where it is acquired. It means that the sensing vectors must be as different as possible from the atoms of $\mathbf{\Phi}$.
%

\subsection{2D-1D inverse problem formulation}
\label{ssec:invPrb}
As indicated in the subsection~\ref{ssec:CS}, the interferometry imaging problem constitutes an ill-posed inpainting problem which can be described mathematically in Eq.~\eqref{eq:visibilities}:
\begin{equation}
\mathbf{V} = \mathbf{M}\mathbf{F}\x + \mathbf{N}
\label{eq:visibilities}
\end{equation}
with $\mathbf{V}$, the measured visibility vector, $\mathbf{M}$ the sampling mask which accounts for incomplete sampling in the Fourier space, $\mathbf{F}$ the FT operator, $\x$ the sky, and $\mathbf{N}$ the noise. The sky $\x$, expressed in the direct space, is a real quantity while the noise $\mathbf{N}$ is complex as it alters both amplitude and phase of the visibility measurements.


By extension, the application to transient imaging leads to recast the entities of Eq. \eqref{eq:visibilities} as time-dependent entities. In that scope, the data model of the sky $\x$, containing a transient source, will be a cube. At a given frequency, two dimensions are associated with the spatial information and the third dimension describe its time dependence. In the classical 2D case, the masking operator $\M$ is a given, depending on the frequency and time integration. In our case, we have to account for its time-dependence as we considered a ground-based radio interferometer that rotates with Earth (leading to the apparent motion of the sky). $\M$  will sample different region of the sky FT with time which is a cumulative effect to the variation of the sampled FT of the varying sky. The imaging of a transient radio sky, can be regarded as a 2D-1D spatial-temporal image reconstruction problem.

\begin{figure}[htbp]
\begin{center}
\includegraphics[width=0.60\linewidth]{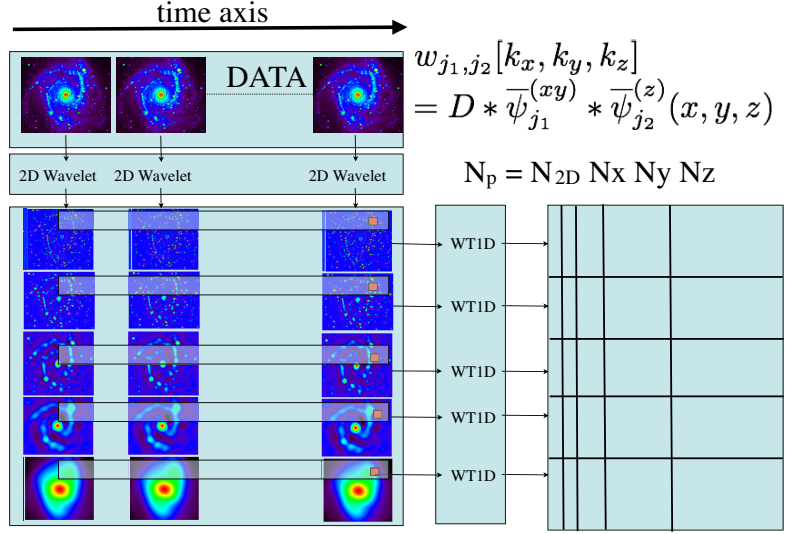}
\caption{Illustration of 2D-1D decomposition: For a cube of size $N_x \times N_y \times N_z$, the total number of coefficients will be $N_{2D} \times N_x \times N_y \times N_z$ where $N_{2D}$ is the 2D decomposition scale.}
\label{2d1d-scheme}
\end{center}
\end{figure}

To comply with the CS framework, the choice of the corresponding 2D-1D sparse representation $\Phi$ is critical for our problem. Fortunately, as the temporal information is not correlated to the spatial information, we can separate the 2D-spatial dictionary and 1D-temporal dictionary rather than looking for a 3D dictionary. Therefore, as described in~\citep{starckfermi}, an ideal wavelet function would be $\psi(x,y,t)=\psi^{(xy)}(x,y)\psi^{(t)}(t)$ where the space (xy) and time (t) are independent. As in \cite{Garsden_2015}, we selected the 2D starlets \citep{starlet} which have proven to be adapted to astronomical sources. For the 1D temporal transform $\psi^{(t)}$, we used decimated wavelet functions such as Haar or biorthogonal CDF 9/7 wavelets, depending on the temporal profile of the transient. The whole 2D-1D decomposition scheme is illustrated in Fig.~\ref{2d1d-scheme}. Firstly, assuming a cube of size $N_x \times N_y \times N_z$ where $N_z$ denotes the number of time frames, the starlet transform is operated on each time frame, yielding $N_{2D}$ spatial scales for each frame. Secondly, For each spatial scale, the temporal 1D transform is performed on each pixel column in the temporal direction of each scale. The 1D transform is decimated and will not increase the size of coefficients in time. Thus, we obtain a 2D-1D coefficient set of size $N_{2D} \times N_x \times N_y \times N_z$.

Given this 2D-1D dictionary $\Phi$, the minimization problem can be formulated from Eq.~\eqref{eq:visibilities} in the analysis framework:
\begin{equation}
\mathbf{min}~||\mathbf{\Phi}^t\x||_1 ~~~\mathit{s.t.}~~~ ||\V-\M\F\x||_2^2 < \epsilon,
\label{eq:min}
\end{equation}
where $\epsilon$ denotes the error radius. The objective minimization function takes the form of $||\mathbf{\Phi}^t\cdot||_1$ where the $\ell_1$-norm (summation of absolute values of coefficients) is used to reinforce the sparsity of the solution and ensure the convexity of the problem. However, the $\ell_1$-norm involves a soft-thresholding operator which induce bias is solutions~\citep{starlet}. This is particularly unsuitable for scientific data analysis involving photometry. According to ~\cite{Candes2007}, the reweighted $\ell_1$ scheme is one way to handle this issue. To address this issue, we adopted a reweighted $\ell_1$ scheme \citep{Candes2007}, by defining a weighting vector $\mathbf{W}$. In addition, as the source photometry is always assumed positive, we impose a positivity constraint as well. Thus, our convex minimization problem can be formulated as:
\begin{equation}
\underset{\x}{\mathbf{min}}~||\V-\M\F\x||_2^2+||\W\odot\bm{\lambda}\odot\mathbf{\Phi}^t\x||_1+\mathit{i}_{\mathbbm{C}^+}(\x),
\label{eq:anaPos}
\end{equation}
where $\bm{\lambda}$, a scale-dependent vector, is a Lagrangian parameter which depends implicitly on $\epsilon$ of the data fidelity in~\eqref{eq:min}, the operator $\odot$ denotes the element-by-element multiplication and $\mathit{i}_{\mathbbm{C}^+}(x)$ is the indicator function (which is 0 if $x$ belongs to $\mathbbm{C}^+$, $+\infty$ otherwise)


As $\bm{\lambda}$ is not explicitly related to the error radius $\epsilon$, the mathematical relation between $\bm{\lambda}$ and $\epsilon$ is not easy to find out. However, since the real dataset $\V$ is noisy and makes the sparsity constraint decline in the 2D-1D decomposition space, $\bm{\lambda}$ is closely related to the statistical distribution of decomposition coefficients. Thus, the study of the statistical distribution is important. In practice, several noise driven strategies are available to estimate the statistical distribution. One of the strategies is noise-driven strategy from the residual, which is used hereof. As we will see in section \ref{ssec:algo}, the residual is obtained by $\mathbf{R}^{(n)}=\V-\M\F\x^{(n)}$ in the n-th iteration. Consequently, the statistical distribution $\bm{\alpha}^{\mathbf{R}^{(n)}}=\mathbf{\Phi}^t\F^t\M^t\mathbf{R}^{(n)}$, and $\bm{\sigma}$ is accessible by the reliable noise estimator MAD (Median of the Absolute Deviation). Then, $\bm{\lambda}=k\bm{\sigma}$ where $k$ defines the level of the significant coefficients which lie within the band $k\sigma$ of the Gaussian distribution.

\subsection{Algorithms}
\label{ssec:algo}
According to~\cite{Candes2007}, the reweighted $\ell_1$ scheme is applied as follows:
\begin{itemize}
\item[1.] Set the iteration count $\mathit{n}=0$ and initialize $\W^{(0)}=\mathbf{1}$.
\item[2.] Solve the minimization problem~\eqref{eq:anaPos} yielding a solution $\x^{(n)}$, and $\bm{\alpha}^{(n)}$ is obtained by $\bm{\alpha}^{(n)}=\mathbf{\Phi}^t\x^{(n)}$.
\item[3.] Update the weights (see later on).
\item[4.] Terminate on convergence or when reaching the maximum number of iterations $N_{\text{max}}$. Otherwise, increment $n$ and go to step 2.
\end{itemize}
First, a biased solution $\x$ is obtained by the non-reweighted convex optimization, then a weighting step is performed using the following weighting strategy: if $|\alpha_{i,j}| \geq k'\sigma_j$ then $w_{i,j}$=$k'\sigma_j/|\alpha_{i,j}|$, else, $w_{i,j}$=1 (operation later described as function $f(|\alpha_{i,j}|)$.
We update the weights for each entity $i$ at scale $j$, $\sigma_j$ is the noise standard deviation at scale $j$. $k'$ acts as a reweight level in scale $j$. 
Then, we subsequently apply the proximal theory to solve the minimization problem~\eqref{eq:anaPos} with the Condat-V\~{u} splitting method (CVSM - ~\cite{condat2013,vu2013splitting}). The CVSM introduces a primal-dual pair $(x,u)$ to solve the convex optimization problem~\eqref{eq:anaPos} using forward-backward algorithm. Thus, in summary, the adapted CVSM with reweighted scheme is presented in Algo~\ref{algo:vu_ana}, where the parameters $\tau$ and $\eta$ respect the convergence condition such as $1-\tau\eta||\Phi||^2>\tau ||\M\F||^2/2$, and $\mu$ is a relaxation parameter used to accelerate the algorithm. If $\mu=1$, the algorithm is in the unrelaxed case or no acceleration case. In the Algo~\ref{algo:vu_ana}, the line 3 can be considered as the forward step to converge the non-negative solution from the residual $\mathbf{R}=\V-\M\F\x$, while the line 4 can be regarded as the backward step to enforce the sparsity constraint by using the soft-thresholding proximity ($\ST$).

\begin{algorithm}[ht!]
\SetAlgoLined
\caption{Analysis reconstruction using CVSM\label{algo:vu_ana}}
\Indm
\KwData{Visibility $\V$; Mask $\M$}
\KwResult{Reconstructed image $\x$}{
}
\Indp
\BlankLine
Initialize $(\x^{(0)},\mathbf{u}^{(0)}),\W^{(0)}=\mathbf{1},\tau>0,\eta>0,\mu\in\left]0,1\right]$\;
\For{$n=0$ \KwTo $N_\text{max}-1$}{
$\mathbf{p}^{(n+1)}=\proj_{\mathbbm{C}^+}(\x^{(n)}-\tau\Phi \mathbf{u}^{(n)}+\tau (\M\F)^*(\V-\M\F\x^{(n)}))$\;
$\mathbf{q}^{(n+1)}=(\mathbf{Id}-\ST_{\bm{\lambda}\odot\W})(\mathbf{u}^{(n)}+\eta\Phi^T(2\mathbf{p}^{(n+1)}-\x^{(n)}))$\;
$(\x^{(n+1)},\mathbf{u}^{(n+1)})=\mu(\mathbf{p}^{(n+1)},\mathbf{q}^{(n+1)})+(1-\mu)(\x^{(n)},\mathbf{u}^{(n)})$\;
$\bm{\alpha}^{(n)}=\mathbf{\Phi}^t \x^{(n)}$\;
Update $\W$ by $w^{(n+1)}_{i,j}=f(|\alpha^{(n)}_{i,j}|)$\;
}
\Return $\x^{(N_\text{max})}$
\end{algorithm}

\section{Experiments}
\label{sec:experiment}
We simulated a sky model of size 32$\times$32$\times$64, i.e. 32$\times$32 pixels on image plane and 64 2-min frames on time. The time-dependent sky model is constituted of a control steady source at the center of the field and a transient source with a gaussian light curve (FWHM = 20 min located at time slice T=24). Both sources have the same peak flux density of 10 arbitrary unit in the sky model. Then, to do the realistic simulation, we generated a 2-hour Fourier sampling mask cube by using an uniform random distribution of 20 antennas observing at the zenith at the latitude of the Nanay Radio Observatory. 

\begin{figure*}[th!]
\begin{center}
\includegraphics[width=0.9\linewidth]{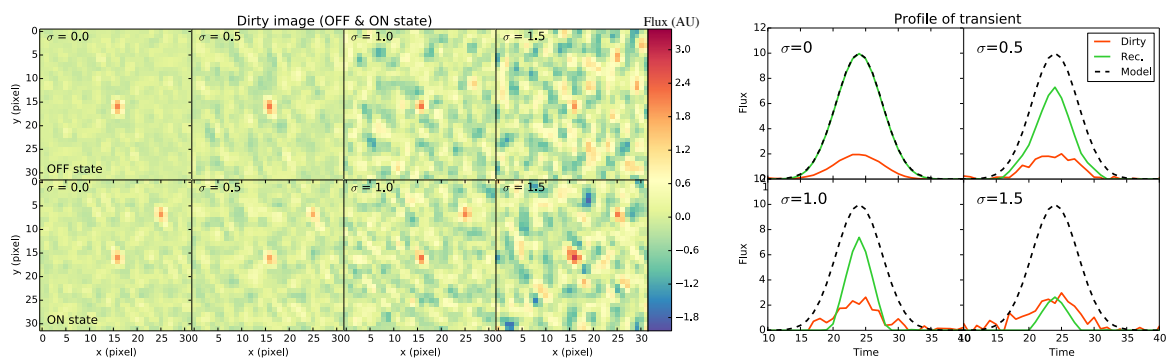}
\caption{(left) Dirty images from the benchmark data cube containing two point sources: a central steady source (x=15,y=15) and a transient source (x=24,y=6). First row corresponds to the OFF state (T=10) and second row to the ON state (T=25) at the maximum of the transient. Each column corresponds to various level of additive gaussian noise with $\sigma=0, 0.5, 1.0, 1.5$ arbitrary flux units. (right) Time profiles at the spatial location of the transient source from the sky model (dash line), the dirty cube (red) and the reconstructed cube (green), for various levels of additive gaussian noise.}
\label{dirtyrecons}
\end{center}
\end{figure*}

To generate the observed visibilities in terms of noise levels, we take the FT of the sky model and apply the mask cube, and then add white gaussian noise with various magnitude ($\sigma = 0.0, 0.5, 1.0, 1.5$ flux unit) on the complex visibilities. Then, by FT inversion, we can obtain the characteristic dirty cube for each case: Fig. \ref{dirtyrecons} (left) shows the transient ``OFF'' state (first row) and its ``ON'' state (second row) of the dirty cube. We notice that when the noise level is high, the transient source can be confused with background artifacts.

Figure \ref{dirtyrecons} (right) illustrates the light curve of the transient source central pixel from the sky model cube (dashed curve), the dirty cube (red curve) and the reconstructed cube obtained by the Condat-V\~u algorithm (green curve) described in Sect. \ref{ssec:algo}. We can conclude that with no additional noise (but with the sampling noise due to missing data), our 2D-1D CS method gives a perfect profile reconstruction, with flux unit relative error $\sim$10$^{-5}$. As the noise level increases, the flux density of the transient is more spread around the central pixel, resulting in a bias of the peak. However, by summing the flux of the transient on nearby pixels (over a surface equal to the source size), the profile of the transient is again well recovered. Meanwhile, the steady source (not shown) is also well recovered except in the high noise regime where fluctuations are present. From these preliminary results, it seems that our 2D-1D CS reconstruction method is efficient reconstruct the transient in a noisy dataset. The reconstruction time was not monitored but is subject to further study.

\section{Conclusions}
\label{sec:conclusion}
CS theory offers new tools for solving ill-posed problems such as the imaging problem in radio interferometry. In previous studies, such as \cite{Garsden_2015}, we have shown that the 2D CS method can outperform classical tools used for deconvolution. In this work, we present an extension of this 2D imager, by recovering the 3D data with the third dimension being the temporal information to detect and reconstruct radio transients. We used respectively a 2D and a 1D wavelet dictionaries to perform 2D-1D reconstruction. In addition, we minimize the convex problem using Condat-V\~u splitting method in the proximal theory framework. The preliminary results based on simulated data cubes containing both steady and transient sources show a good potential for transient imaging. 
For next steps, we will compare our method with classical deconvolution methods and develop a ``time-agile'' imager addressed to the next generation of interferometers, such as LOFAR and SKA, to enable the detection of radio transients. 
A featured paper is in preparation, and contains extended tests on various dataset (simulated and real datasets containing transient source).
As the search for known and unknown transient is a emerging field in radio astronomy, the development of such tools may have a strong impact on transient studies.

\begin{acknowledgements}
We acknowledge the financial support from the UnivEarthS Labex program of Sorbonne Paris Cit\'e (ANR-10-LABX-0023 and ANR-11-IDEX-0005-02) and the Physis project (H2020-LEIT-Space-COMPET).
\end{acknowledgements}

\bibliographystyle{aa}  
\bibliography{jiang} 

\end{document}